\documentclass[%
 reprint,
prb,
showkeys
]{revtex4-2}

\usepackage{graphicx}
\usepackage{dcolumn}
\usepackage{bm}
\usepackage{amssymb}
\usepackage{amsmath}
\usepackage{caption}
\usepackage{subcaption}
\usepackage{blindtext}

\begin{document}

\preprint{APS/123-QED}

\title{Visible Wavelength Flatband in a Gallium Phosphide Metasurface}

\author{Christopher Munley\(^1\)}
\author{Arnab Manna\(^1\)}
\author{David Sharp\(^1\)}
\author{Minho Choi\(^2\)}
\author{Hao Nguyen\(^3\)}
\author{Brandi M. Cossairt\(^3\)}
\author{Mo Li\(^{1,2}\)}
\author{Arthur Barnard\(^{1,4}\)}
\author{Arka Majumdar\(^{1,2,*}\)}
\affiliation{%
\(^1\)Department of Physics, University of Washington, Seattle, Washington 98195, USA. \\
\(^2\)Electrical and Computer Engineering, University of Washington, Seattle, Washington 98195, USA\\
\(^3\)Department of Chemistry, University of Washington, Seattle, Washington 98195, USA\\
\(^4\)Materials Science \& Engineering, University of Washington, Seattle, Washington 98195, USA\\
\(^*\)Corresponding Author\\
}%

\date{\today}

\begin{abstract}
	Engineering the dispersion of light in a metasurface allows for controlling the light-matter interaction strength between light confined in the metasurface and materials placed within its near-field. Specifically, engineering a flatband dispersion increases the photonic density of states thereby enhancing the light-matter interaction. Here, we experimentally demonstrate a metasurface with a flat dispersion at visible wavelengths. We designed and fabricated a suspended one-dimensional gallium phosphide metasurface and measured the photonic band structure via energy-momentum spectroscopy, observing a photonic band that is flat over $10^o$ of half-angle at $\sim 580$nm. We integrated cadmium selenide nanoplatelets with the metasurface, and measured coupled photoluminescence into the flatband. Our demonstration of a photonic flatband will enable the possibility of integrating emerging quantum emitters to the metasurface with possible applications in nonlinear image processing, and topological photonics.
\end{abstract}

                              
\maketitle


\section{\label{sec:level1}INTRODUCTION}
The ability to engineer band-topologies via sub-wavelength patterning has opened up new opportunities in nano-photonic structures, including metasurfaces. Among various band-topologies, photonic flatbands are of particular importance, as the photonic density of states increases near the flatband, with potential applications in enhancing nonlinear interactions between photons and creation of “slow light” \cite{Leykam}. Photonic flatbands have already been experimentally realized in diverse systems, including arrays of waveguides, and coupled cavity arrays \cite{Liqin,Mukherjee}. Metasurfaces have also been employed to realize flatband dispersion in an ultra-compact geometry, but most of them have very poor out-coupling to the free-space mode, and thus are not suitable to access via free-space excitation \cite{Xu}. Photonic flatbands accessible via free space excitation have primarily been limited to the terahertz and infrared metasurfaces \cite{Nakata, NguyenDispersion}. With many emerging quantum emitters in the visible wavelengths, a photonic flatband in this wavelength regime opens up opportunities to study interaction with a greater diversity of emitters. However, realizing a visible flatband metasurface is challenging due to the necessity of a material with high refractive index and fabrication complexity coming from very small geometric features.

Here we designed and fabricated a flatband metasurface at $\sim$ 590nm using gallium phosphide (GaP). The choice of GaP is motivated by its high index ($n\approx3.4$ at the wavelength of interest) and negligible loss in the visible wavelength \cite{Melli}. In fact, GaP photonic structures have already been used to enhance emission from atomically thin 2D material and solution processed quantum materials \cite{Wu, Rivoire}. Here, a commercially available chemical vapor deposited GaP thin film on silicon is partially etched to break the vertical symmetry and then released from the silicon to create a suspended metasurface. We probe the flatband in reflection via energy-momentum spectroscopy, demonstrating a band which is flat over $~10^o$ of half-angle with a measured quality factor estimated to be $1500$, corresponding to a linewidth of $\sim 0.5nm$. Finally, we integrate cadmium selenide (CdSe) nanoplatelets (NPLs) on top of the GaP metasurface and demonstrate flatband-coupled photoluminescence (PL).

\begin{figure}[t]
\centering
\includegraphics[width=\linewidth]{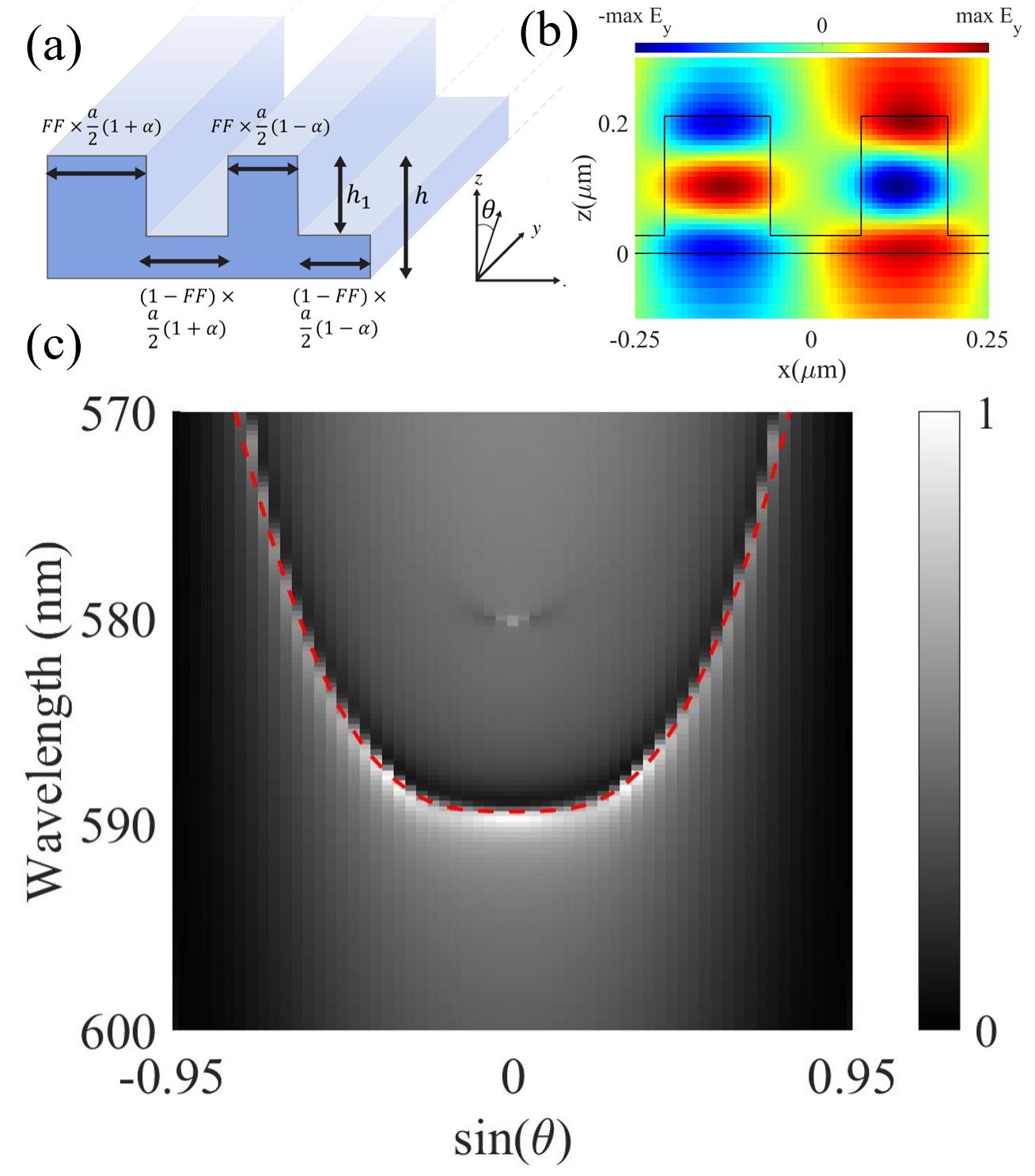}
\caption{a) Schematic of the cross-section of the metasurface showing the period of the metasurface $a$, a period-doubling perturbation $\alpha$, the percentage of the lattice filled with dielectric, called the fill factor $FF$, and the total and partially etched heights ($h$ and $h_1$).We measure the in-plane dispersion as the angle of reflection varies along the y-z plane perpendicular to the direction of periodicity. b) Simulated y-direction electric field of one period of the metasurface at the resonance of the flatband with normally incident light. c) Simulated photonic band structure shows a flat band near $590$nm. The red-dashed line indicates the resonant wavelength determined at each angle by fitting with a Fano resonance.}
\label{fig:Fig. 1}
\end{figure}

\section{\label{sec:level2}DEVICE DESIGN}
\begin{figure}[b]
\centering
\includegraphics[width=\linewidth]{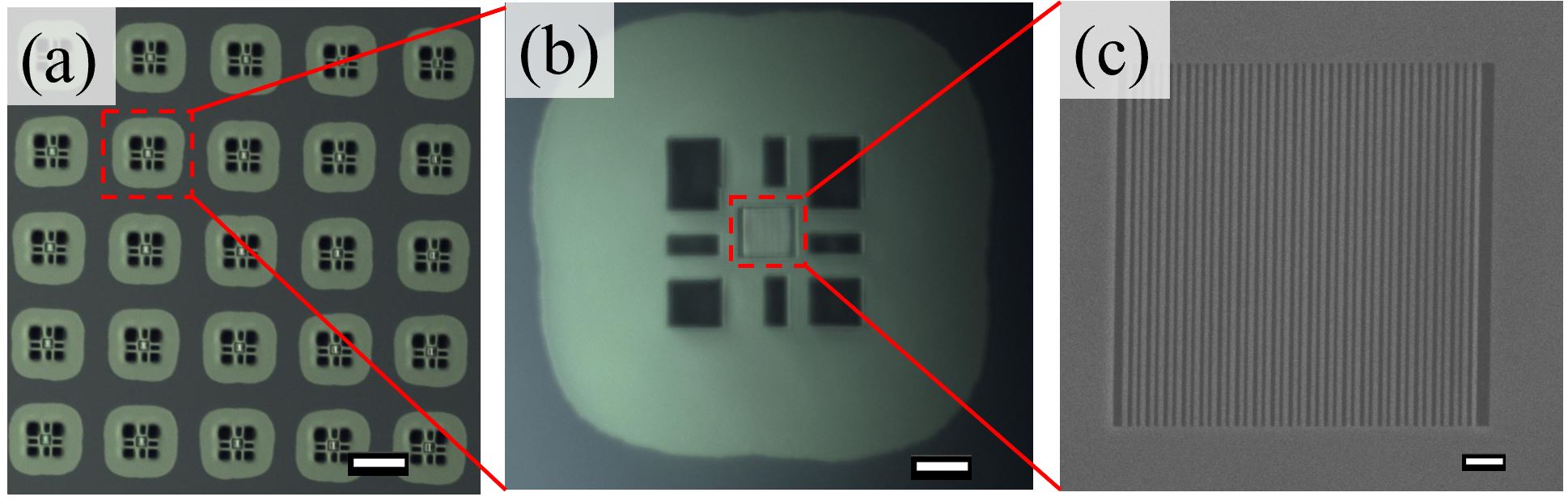}
\caption{a) Optical microscope image of an array of fabricated metasurfaces; scale bar 50$\mu$m. The halo around each structure comes from the membrane being floating. b) Optical microscope image of an individual flatband metasurface; scale bar 10$\mu$m. c) Scanning electron microscope image of the metasurface structures; scale bar 1$\mu$m.}
\label{fig:Fig. 2}
\end{figure}

We designed the flatband metasurface by using rigorous coupled wave analysis (RCWA) in an S4 package \cite{Liu} integrated with Lumerical's Finite-difference time-domain (FDTD) simulation. Our design utilizes vertical symmetry breaking in a dimerized high contrast grating \cite{Zeng} to control band topology while maintaining a mode with a long lifetime (see Fig. ~\ref{fig:Fig. 1}). The vertical symmetry of the metasurface is broken via partial etching of the GaP film. This couples the even and odd modes of a standard periodic grating, and flatband is realized through  the interference of these even and odd modes \cite{NguyenDispersion}.  Additionally, the periodicity of the grating is perturbed by a period-doubling asymmetry, $\alpha$, allowing for efficient free space coupling to bands \cite{Overvig}.  While the period $a$ determines the wavelength of resonance, the fill factor $FF$ and the partial etching depth $h1$ control the dispersion of the band.  While all these parameters are free, all fabricated metasurfaces on the same chip will share the same etch depth, while lateral free parameters like fill factor and period can vary from device to device. We note that the designed flatband in this paper is due to the fine-tuning of parameters, and thus should be considered an``accidental" flatband \cite{Leykam}. Hence, the flat dispersion can be destroyed by any fabrication imperfection. To mitigate this, we fabricate an array of devices, altering the fill factor and period for each device, ensuring that even in the presence of fabrication imperfections such as over and under-etching of the partial etch depth, we will be able to find a device with the designed photonic flatband.  In our device A 210nm thick GaP layer hosts the flatband metasurface, suspended with support structures over the silicon substrate (see Fig. ~\ref{fig:Fig. 2}).  To allow for the maximal range of band tuning by adjusting the fill factor, an optimal partial etch depth of 182nm is selected so that a fill factor of 50 $\%$ results in the desired flat band.

The suspended flatband metasurfaces were fabricated using a 210 nm thick GaP membrane on silicon from the commercial vendor NAsP III/V GmbHe grown via chemical vapor deposition (CVD). A piece of the wafer was spin-coated with 400 nm of Zeon ZEP520A resist and then patterned using a 100 kV JEOL JBX6300FX electron-beam lithography tool to create the mask. The chip is then etched in a reactive ion etcher (RIE) using $Cl_2/Ar$ chemistry. The partial etching was accomplished via appropriate etch-time. A second electron-beam step patterned the ZEP resist to make a mask for etching trenches down to the silicon substrate and the pattern was again transferred to the GaP film using RIE. Finally, the silicon underneath the GaP membrane is removed using vapor phase Xenon difluoride $(XeF_2)$ which delivers a stiction-free and residue-free etching.

\begin{figure*}[t]
\centering
\includegraphics[width=\linewidth]{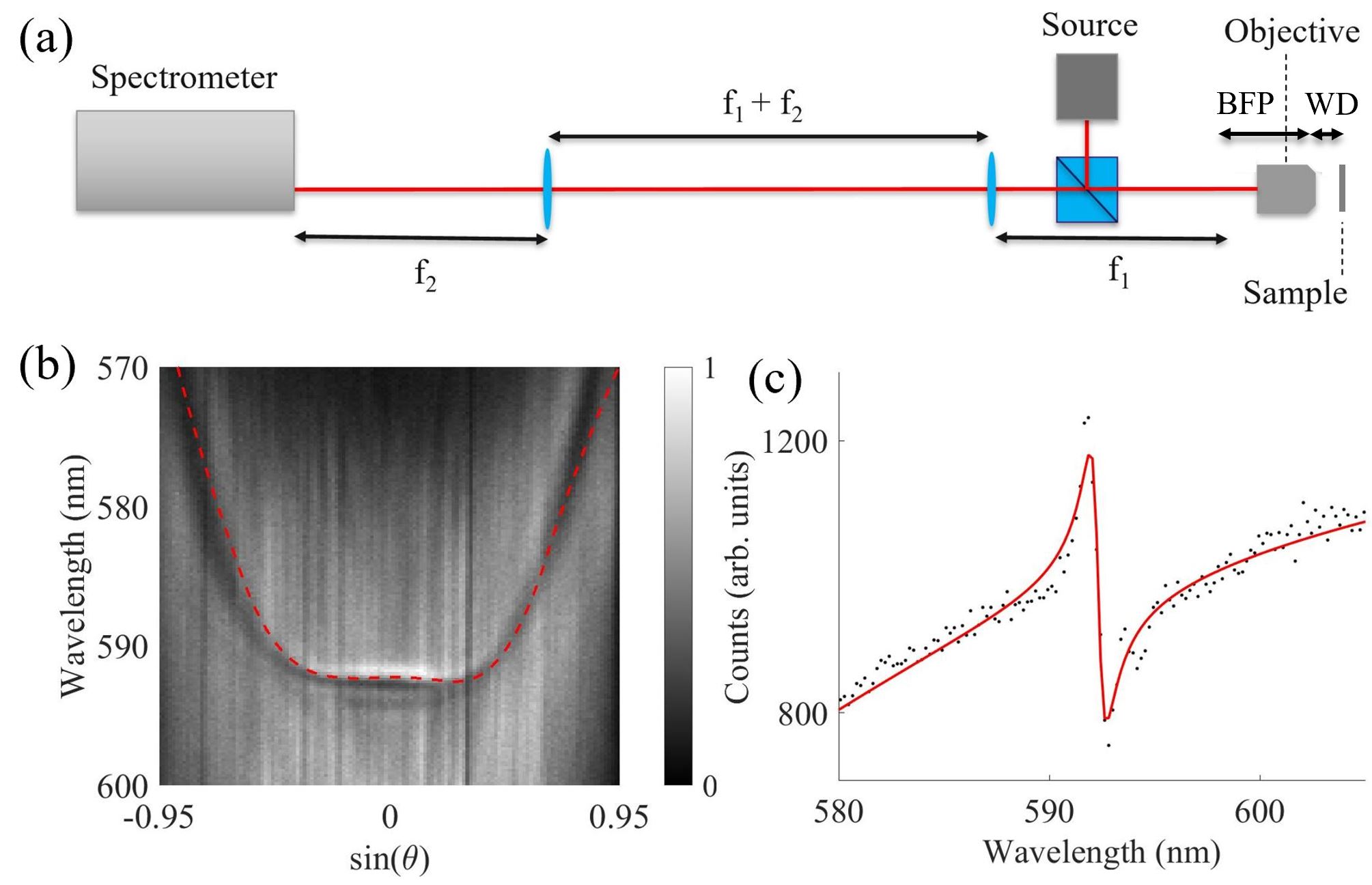}
\caption{a) Schematic of the energy-momentum spectroscopy setup. Working distance of the objective, $WD=.35mm$, back focal distance, $BFP=8mm$, and lenses to image the back focal plane onto the spectrometer with focal distances $f_1=75mm$ and $f_2=150mm$. b) Measured band structure of the flatband device: red-dashed line shows the resonance frequencies obtained via fitting a Fano resonance to the spectrum at each angle of reflection. We clearly observe a flatband dispersion. c) Spectrum of the flatband resonance at the $\Gamma-$point $(\theta=0)$ in reflection normal to the plane of the metasurface; red line shows the fit of a Fano lineshape to the resonance.}
\label{fig:Fig. 3}
\end{figure*}

\section{\label{sec:level3}MEASUREMENT OF BAND STRUCTURE}
We probe the fabricated metasurfaces via energy-momentum (E-K) spectroscopy, employing a spectrometer in combination with a Fourier lens relay to directly image the angle-resolved emission of a sample. A 4f relay is used to image the back focal plane of the objective onto the entrance slit of the spectrometer (see Fig. ~\ref{fig:Fig. 3}). The image of the back focal plane is aligned to the spectrometer slit so that only the emission along the y-axis of the metasurface is collected in the spectrometer. The objective used has a numerical aperture of 0.95, equivalent to greater than $70^o$ of half-angle. This wide angular range allows for the identification and resolution of the angular span of the designed flatband.  For reflection measurements, illumination is provided by a collimated fiber-coupled broadband light from a stabilized tungsten-halogen source (Thorlabs SLS301). The properties of the flatband are extracted directly from this measurement by fitting to determine the quality factor and angular extent of the flatband.

The interference between transmitted light and the resonant band of the metasurface results in a Fano lineshape, when the spectrum of the band is measured at each emission angle. Fitting a Fano resonance to each spectrum, we extract the resonant wavelength and quality factor for the flat band at each angle of emission along the y-axis of the metasurface (see Fig. ~\ref{fig:Fig. 3}).  At the $\Gamma$-point $(\theta=0)$ of the photonic band structure, the quality factor is measured to be $\sim 1500$, across a flatband range of $10^o$ of half-angle.  

\begin{figure}[h]
\centering
\includegraphics[width=\linewidth]{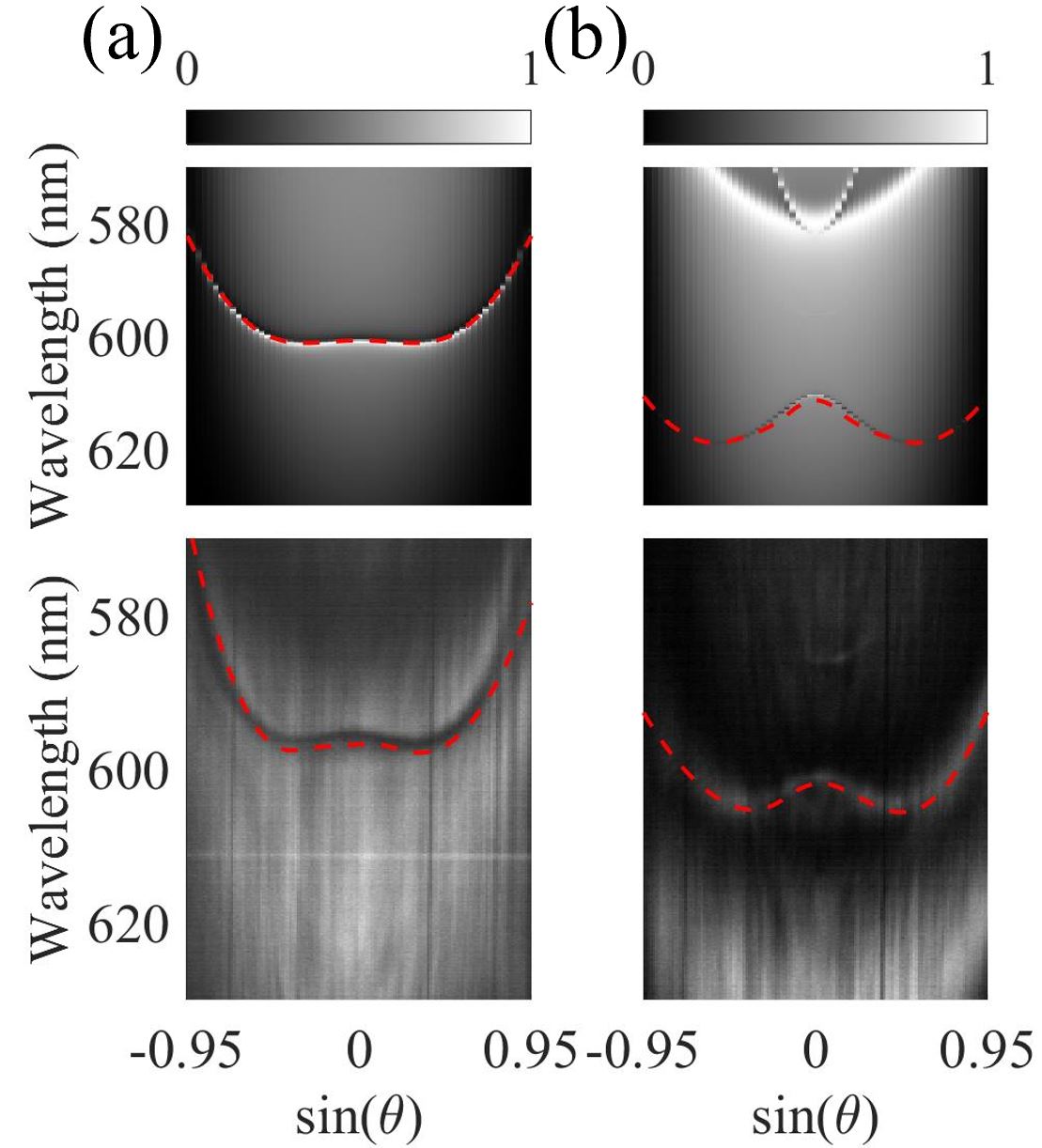}
\caption{a) Measured and simulated near flatband as fill factor is increased and the band bends away from a flatband red-dashed lines indicate a fit of Fano resonances for each angle of emission b) Measured and simulated multivalley W-band dispersion as fill factor is increased even more with resonances shown by red-dashed lines.}
\label{fig:Fig. 4}
\end{figure}

Creation of a flatband is one main application for dispersion engineering, but alternate band shapes have shown promise for realizing novel polaritonic interactions \cite{Pickup}.  As a product of our fabricated array of metasurfaces with different fill factors, we also realize the bending of a flatband into a near flatband and a multi-valley W-band dispersion (see Fig. ~\ref{fig:Fig. 4}).  These multi-valley states have been explored as platforms for valley-polariton systems and exotic phenomena in Bose-Einstein Condensates \cite{SunMultivalley}.  

\section{\label{sec:level4}INTEGRATION WITH CdSe NPLs}
While identification of the visible flatband in reflection validates the metasurface design, to establish the utility of enhancing light-matter interaction, near-field coupling to material is desired. CdSe NPLs, quasi two-dimensional nanocrystals with quantum confined excitons, are particularly promising as emitters due to their narrow inhomogeneous linewidth and ease of integration as a colloidal suspension. Researchers have already shown room temperature strong coupling of NPLs with arrays of plasmonic holes \cite{Pelton,Winkler}.  By controlling the number of monolayers in a NPL, the emission wavelength can be precisely tuned, making them attractive for light emitting devices and polaritonics \cite{Qu, Qiu}. Specifically, six monolayer thick CdSe NPLs have a sharp excitonic resonance at 585nm, just off resonance to the photonic flatband \cite{Cho}. By dropcasting a CdSe NPL solution on the metasurface, we can integrate the emitters without straining the mechanically-fragile floating membrane.

\begin{figure}[b]
\centering
\includegraphics[width=\linewidth]{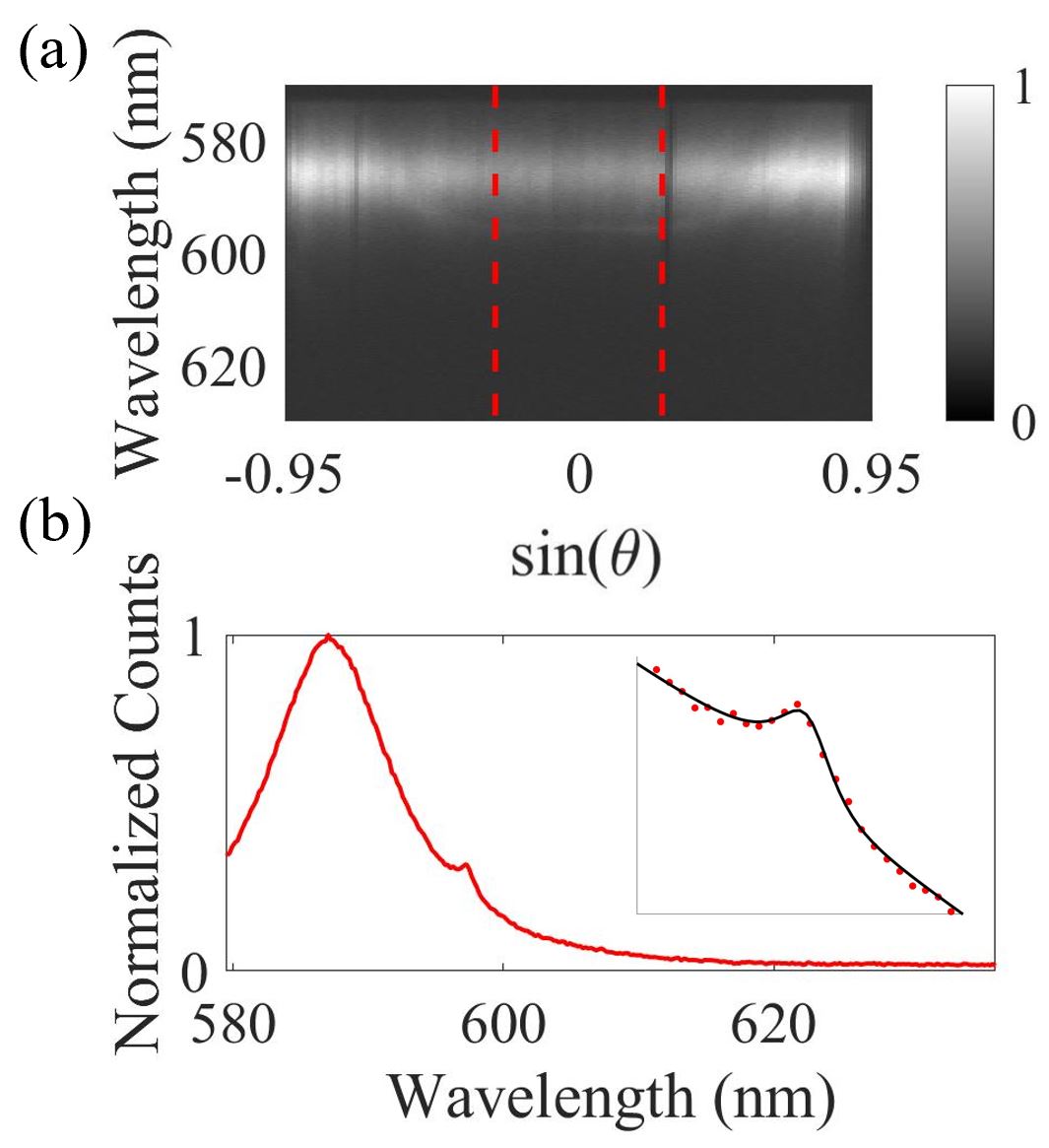}
\caption{a) Measured PL from NPLs coupled to a GaP metasurface showing standard PL extending across all angles of emission as well as the photonic flatband mode enhancing PL.  Red dashed lines highlight the angular extent of the flatband.b) Integrated PL over the flatband as bounded by red-dashed line in (a) showing the main NPL PL peak and the smaller peak due to coupling with the GaP measurface. Inset: the flatband PL peak fit with as a Fano resonance, resulting in a measured quality factor of $Q=1100$.}
\label{fig:Fig. 5}
\end{figure}

Once integrated, the emission of NPL PL into the flatband mode can be verified with above-band excitation via a 532nm laser and energy-momentum spectroscopy.  While a randomly-oriented collection of NPLs will show isotropic emission at its resonant wavelength, emission into the photonic flatband shows enhanced PL outside of the main excitonic peak.  We realize both of these features experimentally, showing the standard NPL PL centered on 587nm and flatband-enhanced PL near 597nm (see Fig. ~\ref{fig:Fig. 5}).  Unlike a photonic crystal defect resonator or traditional grating, the enhancement is not limited to a single location or angle of emission.  Integration of the PL spectrum over the angles of emission in the flatband shows the main PL peak as well as the enhanced flatband emission maintaining a high quality factor, $Q>1000$.  Thus, our GaP flatband metasurface demonstrates a platform for enhancing the PL of emitters at a selected visible wavelength while maintaining strong emission over a wide range of angles.

\section{\label{sec:level5}CONCLUSION}
Our demonstrated visible flatband metasurface represents a step towards future optical applications of engineering dispersion and light-matter interaction. A flatband metasurface in the visible regime creates possibilities in engineering light-matter interaction with emerging quantum emitters.  Flatband coupled PL from CdSe NPLs illustrates one such example. Transitioning our design to mechanically stable, on-substrate platforms will potentially allow for more extensive material integration. Increasing quality factor, for example by reducing the period-doubling asymmetry parameter, or selection of another visible wavlength emitter may allow for potential room temperature strong coupling or the creation of degenerate polariton states. Finally, expansion of the flatband modes to the whole k-space may lead to free-space-integrated slow light metasurfaces.

\begin{acknowledgments}
This material is based upon work supported by the National Science Foundation. Grant No. DMR-2019444. Part of this work was conducted at the Washington Nanofabrication Facility / Molecular Analysis Facility, a National Nanotechnology Coordinated Infrastructure (NNCI) site at the University of Washington with partial support from the National Science Foundation via awards NNCI-1542101 and NNCI-2025489. The authors thank Lorryn Wilhelm for assistance with figure development.
\end{acknowledgments}

\end{document}